\documentclass{emulateapj}
\usepackage{amssymb}
\usepackage{lscape}
\usepackage{longtable}
\begin{document}
\newcommand{\lya}{Lyman~$\alpha$}
\newcommand{\lyb}{Lyman~$\beta$}
\newcommand{\za}{$z_{\rm abs}$}
\newcommand{\ze}{$z_{\rm em}$}
\newcommand{\cmtwo}{cm$^{-2}$}
\newcommand{\nhi}{$N$(H$^0$)}
\newcommand{\degpoint}{\mbox{$^\circ\mskip-7.0mu.\,$}}
\newcommand{\kms}{\,km~s$^{-1}$}      % note leading thinspace
\newcommand{\minpoint}{\mbox{$'\mskip-4.7mu.\mskip0.8mu$}}
\newcommand{\peryr}{\mbox{$\>\rm yr^{-1}$}}
\newcommand{\secpoint}{\mbox{$''\mskip-7.6mu.\,$}}
\newcommand{\sqdeg}{\mbox{${\rm deg}^2$}}
\newcommand{\squig}{\sim\!\!}
\newcommand{\subsun}{\mbox{$_{\twelvesy\odot}$}}
\newcommand{\et}{{\rm et al.}~}
\newcommand{\msun}{\,{\rm M_\odot}}
\newcommand{\Ha}{\,{\rm H\alpha}}
\newcommand{\Hb}{\,{\rm H\beta}}

\def\ltsima{$\; \buildrel < \over \sim \;$}
\def\simlt{\lower.5ex\hbox{\ltsima}}
\def\gtsima{$\; \buildrel > \over \sim \;$}
\def\simgt{\lower.5ex\hbox{\gtsima}}
\def\arcs{$''~$}
\def\arcm{$'~$}
\def\erf{\mathop{\rm erf}}
\def\erfc{\mathop{\rm erfc}}
\title{EVIDENCE FOR SOLAR METALLICITIES IN MASSIVE STAR-FORMING GALAXIES at $z\simgt 2$\altaffilmark{1}}
\author{\sc Alice E. Shapley} 
\affil{University of California, Berkeley, 601 Campbell Hall, Astronomy Department, Berkeley, CA 94720}
\author{\sc Dawn K. Erb}
\affil{California Institute of Technology, MS 105--24, Pasadena, CA 91125}
\author{\sc Max Pettini}
\affil{Institute of Astronomy, Madingley Road, Cambridge CB3 0HA UK}
\author{\sc Charles C. Steidel}
\affil{California Institute of Technology, MS 105--24, Pasadena, CA 91125}
\author{\sc Kurt L. Adelberger}
\affil{Carnegie Observatories, 813 Santa Barbara Street, Pasadena, CA 91101}

\submitted{Accepted for publication in ApJ}

\shorttitle{EVIDENCE FOR SOLAR METALLICITIES IN MASSIVE STAR-FORMING GALAXIES at $z\simgt 2$}
\shortauthors{SHAPLEY ET AL.}

\altaffiltext{1}{Based, in part, on data obtained at the 
W.M. Keck Observatory, which 
is operated as a scientific partnership among the California Institute of Technology, the
University of California, and NASA, and was made possible by the generous financial
support of the W.M. Keck Foundation.
} 

\begin{abstract}
We present results of near-IR spectroscopic measurements of 
7 star-forming galaxies at $2.1 < z <  2.5$. 
Drawn from a large spectroscopic survey of galaxies
photometrically pre-selected by their $U_nG{\cal R}$ colors
to lie at $z\sim 2$, these galaxies were chosen for their
bright rest-frame optical luminosities ($K_s \leq 20.0$).
Most strikingly,
the majority of the sample of 7 galaxies exhibit [NII]/$\Ha$ nebular
emission line ratios indicative of at least solar 
H~II region metallicities, at a lookback time of 10.5~Gyr. 
The broadband colors of the $K_s$-bright sample
indicate that most have been forming stars for
more than a Gyr at $z\sim 2$, and have already formed stellar 
masses in excess of $10^{11} M_{\odot}$.
The descendants of these galaxies in the local universe are
most likely metal-rich and massive spiral and elliptical galaxies,
while plausible progenitors for them can be found
among the population of $z\sim 3$ Lyman Break Galaxies.
While the $K_s$-bright $z\sim 2$ galaxies appear to be highly 
evolved systems, their large $\Ha$ luminosities and uncorrected
$\Ha$ star-formation rates of $24 - 60 M_{\odot}\mbox{yr}^{-1}$
indicate that active star formation is still ongoing.
The luminous UV-selected objects presented here
comprise more than half of the high-redshift ($z>1.5$) tails
of current $K$-band-selected samples such as the K20 and Gemini Deep
Deep surveys.
\end{abstract}

\section{Introduction}
\label{sec:intro}

Rest-frame optical emission lines from extragalactic H~II regions provide an
important diagnostic of galactic chemical evolution and therefore
serve as an independent gauge of the history of star formation 
in the universe. Furthermore, 
the chemical evolution of galaxies and of the IGM are
interconnected processes
since feedback from star formation
in galaxies may be responsible for polluting the IGM
with heavy elements over a wide range of redshifts
\citep{heckman2000,adelbergeret2003a}.
Now, with the advent of sensitive near-infrared spectrographs
on $8-10$~m class telescopes, it is possible to measure
rest-frame optical nebular emission lines out to $z\geq 3$,
and therefore to trace galactic chemical evolution
over the bulk of cosmic history.

In the local universe, star-forming spiral and irregular galaxies
display a strong correlation between their H~II region oxygen
abundances and optical luminosities, according to which more 
luminous and massive galaxies exhibit a higher 
degree of metal enrichment 
\citep{garnett1987,garnett2002,zaritsky1994,lamareille2004,tremonti2004}.
An analogous correlation is observed
in elliptical galaxies, where chemical abundances are measured
from the strengths of stellar magnesium and iron absorption
line indices \citep{brodie1991}. As we move to higher redshifts, the
metallicity-luminosity relationship may evolve
in both zero point and slope \citep{maier2004}.
Using a sample of 64 star-forming galaxies at $0.26 < z < 0.82$ 
drawn from the Deep Extragalactic Evolutionary Probe Groth 
Strip Survey (DGSS), \citet{kobulnicky2003} demonstrate that galaxies
at $0.6 < z < 0.8$ are $1-3$ magnitudes more luminous 
than local galaxies of similar metallicity.
In contrast, \citet{lilly2003} find that a majority
of a sample of 66 star-forming galaxies at
$0.47 < z < 0.92$ drawn from the Canada-France Redshift
Survey (CFRS) have similar metallicities to  galaxies in the local
universe with comparable luminosities.
At much higher redshift, $z> 2.5$, most of the information
about the abundances of galaxies has been derived
from absorption lines in Damped Lyman Alpha systems (DLAs)
\citep[e.g.,][]{pettini1997,prochaska2002}. 
However, relating line-of-sight DLA absorption abundances to
abundances measured from global galactic
H~II emission-line spectra is not trivial, at least in part because
the relationship of DLAs themselves to high-redshift UV-selected galaxies has not
been established. There are only limited 
data on the H~II region oxygen abundances of
star-forming galaxies at $z>2.5$. A sample of
six galaxies has been assembled with near-infrared spectroscopic observations
of the strong nebular emission lines that
enable abundance estimates \citep{pettini2001,kobulnicky2000}. While
there are only rough constraints on the abundances
in these galaxies, with $\mbox{(O/H)} \sim 0.1 - 1.0 \mbox{ (O/H})_{\odot}$,
they are seen to be $2-4$ magnitudes more luminous than local galaxies
of similar metallicity.

Until now, there have been virtually no observations of the
H~II region metallicities in star-forming galaxies at $1.4 < z < 2.5$,
even though this critical epoch may host the production of a large fraction
of the heavy elements present in the local universe.
Galaxies in the ``redshift desert'' range
have only recently become accessible for
detailed study due, in large part, to improvements in both near-UV and 
$0.8-1.0 \mu\mbox{m}$ spectroscopic capabilities on $8-10$~m class
telescopes. Employing new technologies, 
several current surveys utilizing complementary
selection techniques contain spectroscopically confirmed
galaxies at these redshifts. The selection criteria
include optical colors \citep{steidel2004,adelberger2004}, 
$K$-band magnitude
\citep{cimatti2002,glazebrook2004}, $J-K$ color \citep{franx2003},
and submillimeter flux \citep{chapman2003}. A complete picture
of the galaxy population at $z \sim 2$ will require an understanding
of how these several selection techniques overlap and complement each other.

As shown by \citet{pp2004}
and \citet{steidel2004}, H~II region metallicities can be measured in redshift
desert galaxies using the ratio of the [NII]~$\lambda 6584$ and $\Ha$ nebular emission
lines. Here we present near-infrared ($K$-band) spectroscopic observations
of [NII] and $\Ha$ emission for a small
sample of UV-selected star-forming galaxies at $2.1 < z < 2.5$.
These galaxies were selected for observation because they are 
luminous in the rest-frame optical
($K_s \leq 20.0$), and therefore application of the
same observational strategy yields much higher signal-to-noise (S/N)
spectra than previous $z\sim 2$ near-infrared spectroscopic
observations of fainter galaxies \citep{erb03}. 
Along with the spectroscopic observations, which probe
the metal content of the galaxies, we also use broad-band
optical/near-infrared colors to estimate their stellar populations
and masses. With such information, it is possible to consider
the build-up of stellar mass in the early universe, and the 
epoch at which the progenitors of massive galaxies
in the local universe were mostly assembled.

The sample of 
$z\sim 2$ galaxies and
near-infrared spectroscopic observations are described in
\S\ref{sec:sample}. In \S\ref{sec:results}, 
the results of the $\Ha$ and [NII] spectroscopic
measurements are presented. Oxygen abundances are derived 
from these results in \S\ref{sec:abundance}. In \S\ref{sec:discussion}, 
the discussion is extended
to the galaxy metallicity-luminosity relationship,
the stellar populations and masses of the luminous
$z\sim 2$ sample, a comparison of this sample with other
recent surveys, and a comment on inferring stellar masses from
population synthesis modeling versus dynamical masses from $\Ha$ line-widths.
Finally, in \S\ref{sec:summary}, we summarize our main conclusions.
For all luminosity, star-formation rate, and stellar mass calculations, 
we adopt a cosmology with $H_0=70\;{\rm km}\;{\rm s}^{-1}\;{\rm Mpc}^{-1}$,
$\Omega_m=0.3$, and $\Omega_{\Lambda}=0.7$.

\begin{figure}
%\plotone{krk.eps}
\plotone{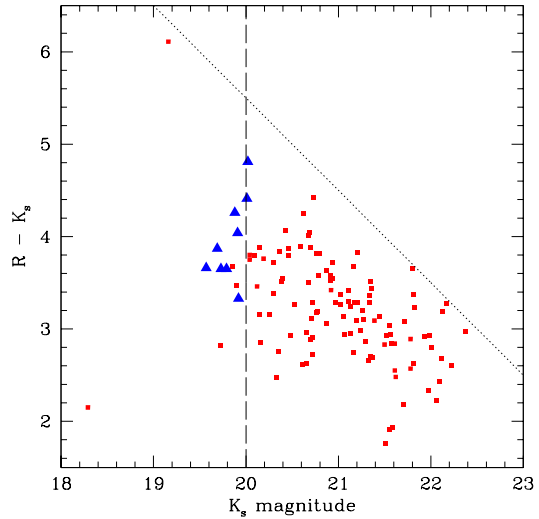}
\caption{The Color-magnitude diagram of 121 UV-selected galaxies at $z> 1$
($\langle z \rangle = 2.28 \pm 0.28$),
in the Q1623 field. ${\cal R}$ and $K_s$ magnitudes are on the AB and Vega systems, respectively.
The dotted diagonal line indicates
the survey magnitude limit of ${\cal R}=25.5$. Only $\sim 10$\%
of the objects at $z> 1$ also have $K_s\leq 20.0$,
indicated by the long-dashed line. The nine $K\leq 20.0$ objects
observed with NIRSPEC in September 2003 are indicated by large blue
triangles. The object in the lower left-hand corner, with
$K_s=18.29$ and ${\cal R} -K_s=2.15$, is a QSO at $z=2.529$.
The median color uncertainty is $\sigma({\cal R}- K_s)=0.25$, though
the 10 $K_s$-selected objects have $\sigma({\cal R}-K_s) \sim 0.10$.
}
\label{fig:krk}
\end{figure}

\section{Sample and Observations}
\label{sec:sample}

\begin{deluxetable*}{llllllccccc}
\tablewidth{0pt}
\tabletypesize{\footnotesize}
\tablecaption{Galaxies Observed With Keck~II/NIRSPEC\label{tab:obs}}
%\rotate
\tablehead{
\colhead{Galaxy} &
\colhead{R.A. (J2000)} &
\colhead{Dec. (J2000)} &
\colhead{$z_{\rm Ly\alpha}$\tablenotemark{a}} &
\colhead{$z_{\rm abs}$\tablenotemark{b}} &
\colhead{$z_{\Ha}$\tablenotemark{c}} &
\colhead{${\cal R}_{AB}$} &
\colhead{$(G-\cal R)_{AB}$} &
\colhead{$(U_n-G)_{AB}$} &
\colhead{${\cal R}_{AB} - K_{s,Vega}$} &
\colhead{Exposure (s)}
}

\startdata

Q1623-BX274 & 16 25 38.202 & 26 45 57.141 & 2.415   & 2.408 & 2.4100 & 23.23 & 0.25 & 0.89 & 3.66 & 3$\times$900 \\
Q1623-MD66  & 16 25 40.392 & 26 50 08.878 & \nodata & 2.111 & 2.1075 & 23.95 & 0.37 & 1.40 & 4.04 & 3$\times$900 \\
Q1623-BX341 & 16 25 43.554 & 26 46 36.942 & \nodata & 2.377 & \nodata & 24.83 & 0.47 & 0.90 & 4.81 & 2$\times$900 \\
Q1623-BX344\tablenotemark{d} & 16 25 43.931 & 26 43 41.977 & \nodata & 2.422 & 2.4224 & 24.42 & 0.39 & 1.25 & 4.41 & 2$\times$900 \\
Q1623-BX453 & 16 25 50.836 & 26 49 31.399 & 2.183   & 2.171 & 2.1816 & 23.38 & 0.48 & 0.99 & 3.65 & 3$\times$900 \\
Q1623-BX513 & 16 25 55.856 & 26 46 50.304 & 2.249   & 2.244 & 2.2473 & 23.25 & 0.26 & 0.68 & 3.33 & 2$\times$900 \\
Q1623-BX528 & 16 25 56.439 & 26 50 15.444 & \nodata & 2.266 & 2.2682 & 23.56 & 0.25 & 0.71 & 3.87 & 4$\times$900 \\
Q1623-BX599 & 16 26 02.545 & 26 45 31.900 & \nodata & 2.329 & 2.3304 & 23.44 & 0.22 & 0.80 & 3.65 & 4$\times$900 \\
Q1623-BX663\tablenotemark{e} & 16 26 04.576 & 26 48 00.202 & 2.435   & \nodata & 2.4333 & 24.14 & 0.24 & 1.02 & 4.26 & 3$\times$900 \\
\enddata
\tablenotetext{a}{Vacuum heliocentric redshift of Ly$\alpha$ emission
  line, when present.}
\tablenotetext{b}{Vacuum heliocentric redshift from rest-frame UV
  interstellar absorption lines.}
\tablenotetext{c}{Vacuum heliocentric redshift of H$\alpha$ emission
  line.}
\tablenotetext{d}{BX344 was observed with a 0\secpoint57 slit,
while all other galaxies were observed with a 0\secpoint76 slit.}
\tablenotetext{e}{The H$\alpha$ emission in BX663 has a two distinct
peaks, one at $z=2.4333$, as listed above, and another at $z=2.4289$.}

\end{deluxetable*}

The objects discussed here were drawn from the survey presented
by \citet{steidel2004}. This survey photometrically identifies
star-forming galaxies at $z\sim 2$ by their locations in 
$U_n-G$ vs. $G-{\cal R}$ color space. Objects satisfying the
optical color criteria presented in \citet{adelberger2004} have
been followed up spectroscopically using the Low Resolution
Imaging Spectrometer on the Keck~I telescope \citep{oke1995}.
To date, 692 galaxies have been spectroscopically confirmed with
$1.4 \leq z \leq 2.5$. The survey has been conducted in seven
separate fields, five of which contain one or more bright background
QSOs at $z \simgt 2.5$ in order to study the evolving relationship
between star-forming galaxies, gas and metals in the
intergalactic medium (IGM). The remaining two pointings are the GOODS-N
and Westphal/Groth fields, both of which have extensive multi-wavelength
datasets that are or will soon be publicly available. In addition
to optical imaging and spectroscopic data, in June and October 2003 we
obtained near-infrared $K_s$-band photometry for galaxies in
three of the survey QSO fields using the Wide Field Infrared
Camera (WIRC) on the Palomar 5.1~m Hale telescope. $K_s$ magnitudes
have been measured for 283 spectroscopically confirmed
galaxies at $z>1.4$, with $\langle z \rangle = 2.25 \pm 0.31$. 
One of these three fields, 
Q1623 (1623+27), contains 167 spectroscopically confirmed 
galaxies at $z>1.4$ in the region covered by WIRC, 
121 of which were detected at $K_s$.

At the redshifts
probed in the sample, $K_s$-band corresponds most closely to rest-frame
optical ${\cal R}$-band, and for $z \geq 1.9$ 
$K_s$-band spectroscopic observations
cover the $\Ha$ and [N~II] nebular emission lines. Using
NIRSPEC \citep{mclean1998} on the Keck~II telescope, we have assembled
a sample of more than 50 $\Ha$ measurements in the $K_s$-band
\citep{erb03, erb04}.
Since the majority of these objects were chosen
primarily for their close proximity 
to a QSO sightline 
\citep[see][for a discussion]{erb03}, the
NIRSPEC sample contains a fairly unbiased view of 
the range of rest-frame UV and optical
properties of galaxies in the \citet{steidel2004} survey.
With this unbiased sample (with respect to photometric properties) 
in place as a control, we used the
$K_s$ photometry to select galaxies for NIRSPEC rest-frame
optical spectroscopic follow-up.
In Figure~\ref{fig:krk}, we present the distribution of $K_s$
magnitudes and ${\cal R}-K_s$ colors for spectroscopically
confirmed galaxies in the Q1623 field.
As shown in the figure,
10\% of the galaxies in the Q1623 field
with WIRC photometry have $K_s \leq 20.0$, (and would be
included in the $K20$ survey \citep{cimatti2002}, 
the high-redshift tail of which
overlaps with the UV-selected sample here under consideration),
while 8\% have ${\cal R}-K_s \geq 4.0$. In order to
measure the rest-frame optical spectra of
objects with bright rest-frame optical luminosities,
we targeted galaxies with $K_s \leq 20.0$. 

The Keck~II/NIRSPEC observations were conducted during the course of
a single observing run spanning 9 -- 13 September 2003. We targeted
a total of nine spectroscopically confirmed $z\sim 2$
objects in the Q1623 field satisfying $K_s \leq 20.0$,
four of which also had ${\cal R}-K_s \geq 4.0$.
All but one of these galaxies satisfy the ``BX'' color criteria presented
in \citet{adelberger2004}, while the remaining galaxy falls
in the adjacent region of $U_nG{\cal R}$ color space occupied
by ``MD'' objects \citep{steidel2003}. The single MD
object in the sample falls on the border between the two color
criteria, lies at the same redshift as the other eight galaxies,
and is therefore treated identically for the remainder of the analysis
and interpretation. We followed the same observational 
procedures described in detail in \citet{erb03}.
In brief, we used the NIRSPEC6 filter, which transmits from 
$1.84-2.63$~$\mu$m. Since the medium-dispersion mode in which
we were observing only allows for $\sim 0.4$~$\mu$m of coverage to fall
on the NIRSPEC detector at once, we set the NIRSPEC cross-disperser angle
to cover $1.9 - 2.3$~$\mu$m, which encompassed $\Ha$ and [NII]
for the redshifts of all of our target galaxies. For all objects
except BX344, we used a 0\secpoint76~$\times$~42'' long-slit,
affording a spectral resolution of $\sim 15$~\AA, whereas for BX344,
we used a 0\secpoint56~$\times$~42'' long-slit, affording $\sim 12$~\AA\
spectral resolution, as measured from the widths of skylines.\footnote{BX344
was the first near-IR-selected object targeted during the observing run,
and we used it to experiment with a narrower slit setup. A
narrower slit reduces the effects of bright sky lines, both due to 
the reduced sky flux in the lines themselves, and the reduced number of 
wavelengths affected by sky flux because of
increased spectral resolution. However,
we found that difficulties in centering the object on the narrower
slit and the resulting variable flux losses outweighed the advantages of 
the narrower slit. Therefore, the remainder of the near-IR selected
objects were observed with the 0\secpoint76 slit.} The same method
of blind-offset object acquisition was used as described in \citet{erb03}. In
most cases, the slit position angle was determined by the attempt to 
fit two galaxies on the slit (usually our primary near-IR-selected
target and a nearby companion with or without a Keck~I/LRIS-confirmed
redshift), though in the case of BX513, which
appeared elongated in the WIRC $K_s$ image, the slit position angle
was chosen to lie along the major axis of the galaxy light.
Targets were observed for 2, 3, or 4 $\times$~900~seconds. We detected
eight out of nine of our target galaxies, and the one undetected galaxy, BX341, was
only observed for $2\times900$~seconds. Conditions were photometric
throughout the run with approximately 0\secpoint5 FWHM seeing
in the NIRSPEC6 band. 
A summary of the observations
including target coordinates, redshifts, optical and near-IR 
photometry, and total exposure times is given in Table~\ref{tab:obs}.
The two-dimensional galaxy spectral images were then reduced, 
extracted to one dimension along
with 1-$\sigma$ error spectra, and flux-calibrated
with observations of A stars, according to the procedures outlined
in detail in \citet{erb03}.

\section{Results}
\label{sec:results}
\begin{figure*}
%\plotone{testha.ps}
\plotone{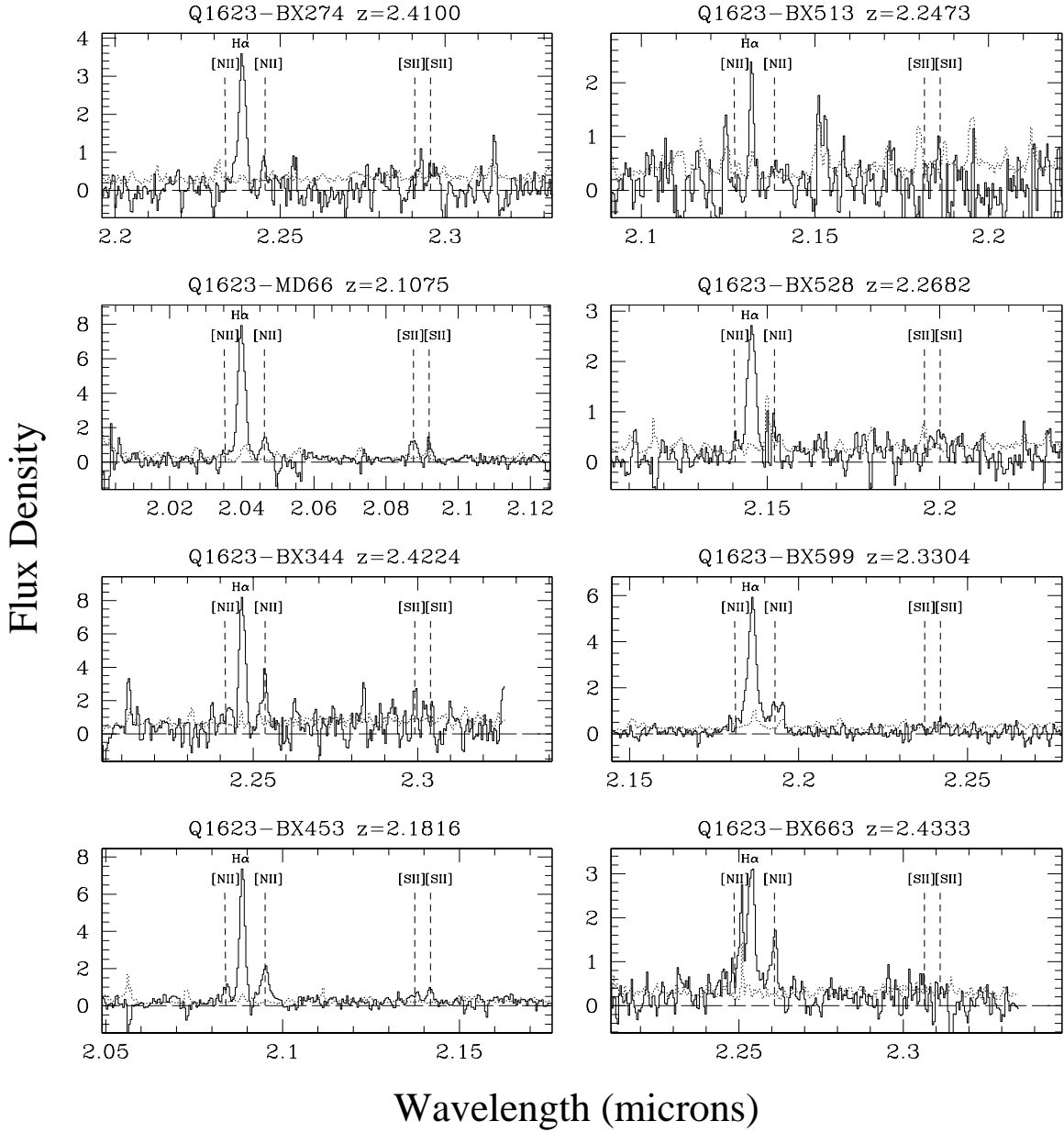}
\caption{Fully-reduced one-dimensional NIRSPEC spectra. The
$\Ha$, [NII], and [SII] lines are marked in each spectrum. The
solid line indicates flux, while the dotted line represents
the error spectrum. The vertical scale indicates flux-density in units of
$10^{-18} \mbox{ erg s}^{-1}\mbox{\AA}^{-1}$. We exclude
BX513 (upper right hand panel)
from all analysis, since its $\Ha$ emission line falls directly on top
of a sky line.
}
\label{fig:halphaspec}
\end{figure*}

The one-dimensional, flux-calibrated NIRSPEC spectra are shown in
Figure~\ref{fig:halphaspec}. $\Ha$ and [NII] $\lambda$~6584
emission are very significantly
detected in all spectra except that of BX513. The redshift
of this object places the $\Ha$ emission line directly on top
of a sky OH emission feature at 2.1318 $\mu$m, 
thus making flux measurements extremely uncertain.
We therefore exclude BX513 from all subsequent analysis, leaving us 
with a robust sample of seven galaxies.\footnote{While [NII]
is not detected in the BX513 spectrum, the limits on [NII]/$\Ha$
are not very restrictive, given the low S/N 
and sky contamination
of the $\Ha$ line.} [SII] emission is also significantly detected 
in the spectra of MD66 and BX453, and marginally
detected for BX274, BX344, BX599, and 
BX528. $\Ha$ and [NII] emission line fluxes were determined by first
fitting a one dimensional Gaussian profile to the higher S/N $\Ha$ feature
to obtain the redshift and FWHM (in wavelength). The $\Ha$ redshift and FWHM
were then used to constrain the fit to the [NII] emission line. This
procedure is based on the assumption that the $\Ha$ and [NII] lines
have exactly the same redshift and FWHM, but that the $\Ha$
line offers a higher S/N estimate of these parameters. Such an
assumption appears to describe all of the spectra well except
that of BX453, in which the [NII] line has broader wings
than the $\Ha$ line, and the fit based on the
$\Ha$ parameters underestimates the [NII] flux by $\sim 30$\%.
\citet{erb04} find an average velocity dispersion of 
$\langle \sigma \rangle =114$~\kms\ for the sample of 61 
UV-selected $z\sim 2$ galaxies
with $\Ha$ measurements from which the current sample is drawn. Five
of the seven $K_s$-selected objects have $\sigma > 114$~\kms,
though two of them, BX453 and BX344, have velocity dispersions
significantly below the \citet{erb04} average.
We list the measured $\Ha$ velocity dispersions, $\Ha$ and [NII] fluxes,
[NII]/$\Ha$ flux ratios, and random statistical uncertainties
on these quantities in Table~\ref{tab:abun}.

Even with relatively short exposure times,
we obtained highly significant detections of $\Ha$ emission
for our target objects. The mean $\Ha$ flux is
$1.34 \times 10^{-16} \mbox{ erg s}^{-1}\mbox{cm}^{-2}$
with a standard deviation among the seven galaxies of 
$0.50 \times 10^{-16} \mbox{ erg s}^{-1}\mbox{cm}^{-2}$;
the corresponding average $\Ha$ star-formation rate
is $47 \pm 15 M_{\odot}\mbox{yr}^{-1}$ using the calibration of 
\citet{kennicutt1998}. These values are 
almost three times larger than the average $\Ha$
flux and star-formation rate of the sample considered in \citet{erb03}. 
The average rest-frame optical continuum luminosity for 
these galaxies is also roughly three times brighter than
that of the \citet{erb03} sample, so the average $\Ha$
equivalents widths of the two samples are very similar.
Even after correcting the broadband
$K_s$ magnitudes for the contribution from
$\Ha$ emission lines, five of seven galaxies in our sample
would still have been selected as $K_s \simlt 20$, whereas the
remaining two galaxies, BX344 and MD66, would have almost
made the cut, with $K_s=20.20$ and
$K_s=20.08$, respectively. In other words,
for five of the seven galaxies in the sample presented here, the $\Ha$
emission line contributes less than 0.15 of the flux in the $K_s$-band. For
the remaining two galaxies, the $\Ha$ line contributes between
0.15 and 0.20 of the $K_s$-band flux. 

\begin{figure*}
%\plotone{2dspec.ps}
%\epsscale{0.7}
\plotone{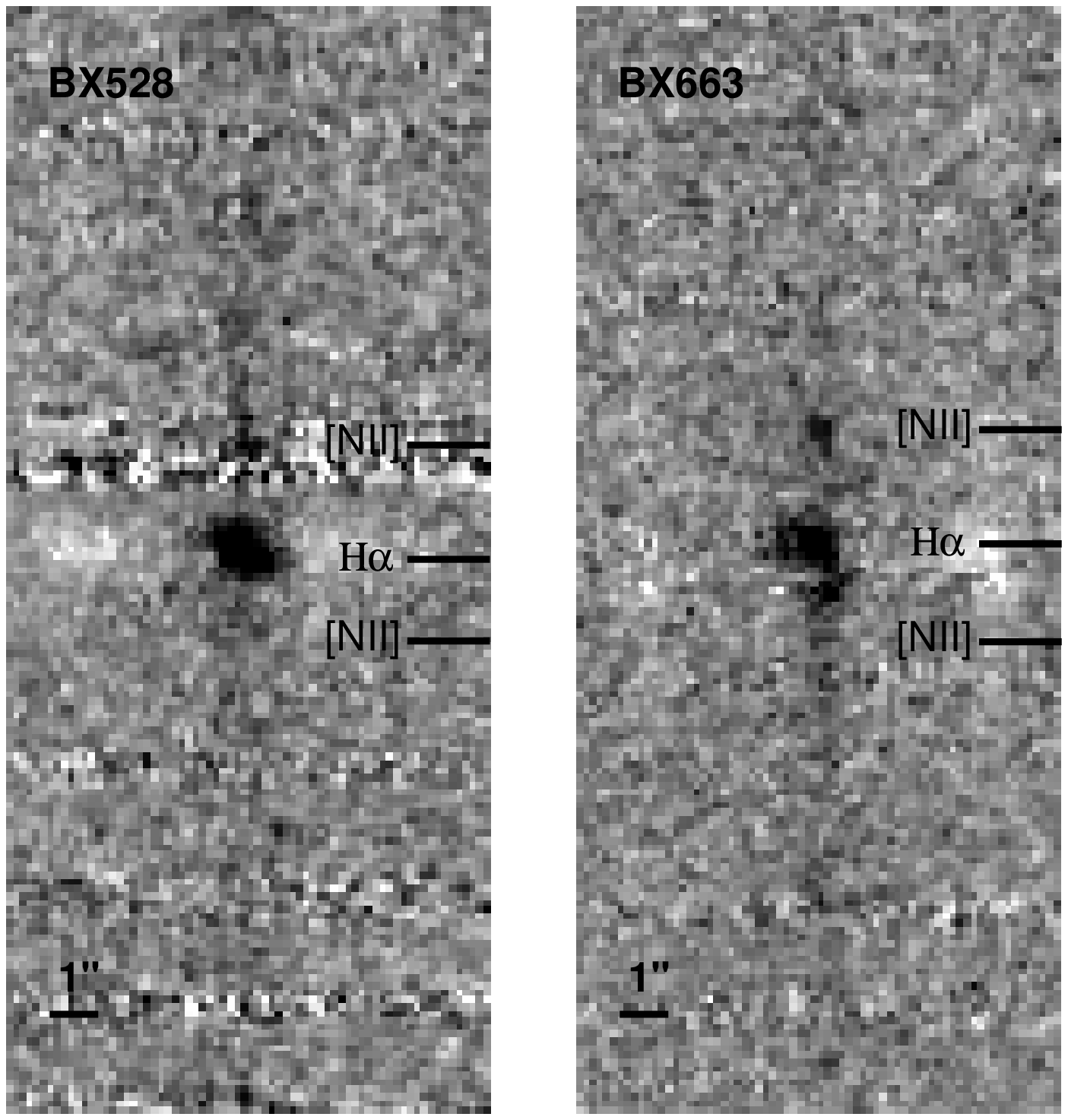}
\caption{Two dimensional spectra for objects with spatially resolved $\Ha$
emission.
Both the spatial scale and positions of $\Ha$ and [NII] emission lines
are marked.
(Left): The spectrum of BX528, which exhibits a tilted $\Ha$ emission
line with a shear of $v_{max} - v_{min} \sim 140$~\kms.
(Right): The spectrum of BX663, which has a complex $\Ha$ emission
structure consisting of two components that are offset by $390$~\kms\
in velocity. The higher-redshift component is spatially extended towards
the left of the continuum, while the lower-redshift component tilts
up and to the right of the continuum. This double spatial structure is
mirrored in the much fainter [NII] emission.
}
%\epsscale{1.0}
\label{fig:ha2d}
\end{figure*}

Most of the galaxies have spatially unresolved $\Ha$ profiles,
with the exception of BX528 and BX663. As shown in Figure~\ref{fig:ha2d},
BX528 has a spatially resolved and tilted $\Ha$ line, even though the
slit position angle was not selected to lie along any 
morphologically-preferred axis. The extent of the
velocity shear in BX528 is $v_{max} - v_{min} \sim 140$~\kms. 
BX663 has a double-peaked $\Ha$ emission line, 
with a velocity separation of $\Delta v = 390$~\kms\
between the two peaks. The higher-redshift $\Ha$ component is 
$\sim 3$ times more luminous than the lower-redshift component
and fairly extended,
with detected flux extending $\sim 2 \arcsec$ along the slit.
While there is a tentative detection of [NII]
for the fainter, lower-redshift $\Ha$ component, we only list the 
[NII]/$\Ha$ ratio for the 
more-significantly detected, higher-redshift component.
The $\Ha$ flux for BX663 listed in Table~\ref{tab:abun} represents the
sum of the two components.

The most striking result, shown in Figure~\ref{fig:halphaspec} and
Table~\ref{tab:abun}, is the high [NII]/$\Ha$ ratios measured
for the $K_s$-selected galaxies. The average ratio is 
$\mbox{[NII]}/\Ha = 0.27$, and the measured values range from
$0.17 - 0.43$. It is relevant to consider the lowest
[NII]/$\Ha$ ratio that could have been measured, given an
$\Ha$ S/N. In the sample of \citet{erb03},
while [NII] is detected for only one object (Q1700-BX691, which
has an above-average $\Ha$ flux), the [NII]
non-detections for the remainder of the sample are 
not very restrictive on the value of [NII]/$\Ha$. Since 
the $\Ha$ emission lines in the \citet{erb03} sample are typically
detected with a significance of $\sim 10$ times 
the random noise, [NII] will only be significantly
detected if it attains $\sim 0.30$ the strength of $\Ha$.
To estimate the typical [NII]/$\Ha$ for the galaxies in \citet{erb03},
we constructed an average composite spectrum from
the individual NIRSPEC spectra, and found
$\langle \mbox{[NII]}/\Ha \rangle = 0.10$.
In the $K_s$-selected sample, the higher significance of the $\Ha$ detections
allows for a more sensitive probe of individual galaxies' [NII]/$\Ha$
ratios. We could have detected $\mbox{[NII]}/\Ha < 0.10$
for BX453, MD66, and BX599, and $\mbox{[NII]}/\Ha = 0.12 - 0.13$
for BX274, BX344, and BX663. BX528 is the only galaxy whose
[NII]/$\Ha$ ratio of $0.19$ is at the limit of what we could have detected.

\begin{deluxetable*}{lccrcccc}
\tablewidth{0pt}
\tabletypesize{\footnotesize}
\tablecaption{$\Ha$ and [NII] Measurements and the Abundance of Oxygen\label{tab:abun}}
%\rotate
\tablehead{
\colhead{Galaxy} &
\colhead{$z_{\Ha}$\tablenotemark{a}} &
\colhead{$\sigma$\tablenotemark{b}} &
\colhead{$F_{\Ha}$\tablenotemark{c}} &
\colhead{$F_{\rm{[NII]}}$\tablenotemark{c}} &
\colhead{$\rm{[NII]}/\Ha$} &
\colhead{$12+\log (O/H)$\tablenotemark{d}} &
\colhead{$12+\log (O/H)$\tablenotemark{e}} \\
\colhead{} &
\colhead{} &
\colhead{(\kms)} &
\colhead{} &
\colhead{} &
\colhead{} &
\colhead{(Pettini \& Pagel)} &
\colhead{(Denicol{\' o})}
}
\startdata

Q1623-BX274 & 2.4100 & $121 \pm 10$ & $9.5 \pm 0.4$ & $1.6 \pm 0.4$ & $0.17 \pm 0.04$ & $8.47 \pm 0.19$ & $8.56 \pm 0.21$ \\
Q1623-MD66  & 2.1075 & $120 \pm 5$ & $19.7 \pm 0.4$ & $3.4 \pm 0.4$ & $0.17 \pm 0.02$ & $8.47 \pm 0.18$ & $8.57 \pm 0.20$ \\
Q1623-BX344 & 2.4224 & $\mbox{  }92 \pm 9$ & $17.1 \pm 0.7$ & $6.2 \pm 0.8$ & $0.36 \pm 0.05$ & $8.65 \pm 0.18$ &  $8.80 \pm 0.20$ \\
Q1623-BX453\tablenotemark{f} & 2.1816 & $61 \pm 6$ & $13.8 \pm 0.3$ & $4.1 \pm 0.3$ & $0.30 \pm 0.02$ & $8.60 \pm 0.18$ & $8.74 \pm 0.20$ \\
Q1623-BX513 & 2.2473 & \nodata\tablenotemark{g} & $3.3 \pm 0.3$ & $<0.9$ & $< 0.30$ & $<8.60$ & $<8.73$ \\
Q1623-BX528 & 2.2682 & $142 \pm 19$ & $7.7 \pm 0.5$ & $1.5 \pm 0.5$ & $0.19 \pm 0.07$ & $8.49 \pm 0.20$ & $8.60 \pm 0.23$ \\
Q1623-BX599 & 2.3304 & $162 \pm 9$ & $18.1 \pm 0.6$ & $4.7 \pm 0.6$ & $0.26 \pm 0.03$ & $8.57 \pm 0.18$ & $8.69 \pm 0.20$ \\
Q1623-BX663 & 2.4333 & $132 \pm 15$ & $16.8 \pm 0.9$\tablenotemark{h} & $3.5 \pm 0.3$\tablenotemark{h} & $0.43 \pm 0.05$\tablenotemark{h} & $8.69 \pm 0.18$ & $8.85 \pm 0.20$ \\
\enddata
\tablenotetext{a}{Vacuum heliocentric redshift of H$\alpha$ emission.}
\tablenotetext{b}{$\Ha$ velocity dispersion obtained by fitting a Gaussian profile to the $\Ha$ line and
deconvolving the effects of instrumental resolution.}
\tablenotetext{c}{ $\:$Line flux and random error in units of $10^{-17}\mbox{ erg s}^{-1}\mbox{ cm}^{-2}$.
While the systematic flux uncertainties are $\sim 25$\%, the uncertainty in the [NII]/$\Ha$
flux-ratio is determined by the random errors in both line fluxes.}
\tablenotetext{d}{Oxygen abundance deduced from the
relationship presented in \citet{pp2004}. For comparison, the most recent estimate
of the solar abundance is $12+\log(\mbox{O/H})_{\odot} = 8.66 \pm 0.05$,
and that of the Orion nebula is $12+\log(\mbox{O/H})_{Orion} = 8.64 \pm 0.06$
\citep{allende2002,asplund2004,esteban1998}.}
\tablenotetext{e}{Oxygen abundance deduced from the relationship presented in \citet{denicolo2002}.}
\tablenotetext{f}{The $\Ha$ and [NII] line fluxes presented are integrals under
Gaussian fits to the lines, whose central wavelengths and FWHM are determined by the parameters of the $\Ha$ line. For
all objects except BX453, the fluxes obtained from the fits to the lines agree with the non-parametric
integrals under the spectra, well within the uncertainties. However, in the case of BX453, the [NII]
flux obtained from the fit is 30\% lower than that obtained by integrating non-parametrically
under the spectrum, due to the larger apparent FWHM of the [NII] line than the $\Ha$ line.
 }
\tablenotetext{g}{The $\Ha$ emission from BX513 falls directly on top of a skyline, preventing a measurement of $\sigma$.}
\tablenotetext{h}{The $\Ha$ line flux listed for BX663 represents the sum of
the two components integrated over the entire extended
region of $\Ha$ emission. The [NII] line flux and
[NII]/$\Ha$ ratio correspond to the
more-significantly detected, higher-redshift component, and
only include flux from the spatial extent common to both transitions.}

\end{deluxetable*}

\section{The Oxygen Abundance}
\label{sec:abundance}

The most robust measurements of chemical abundances
from nebular emission lines are derived from the temperature-sensitive
ratio of an upper- and intermediate-level transition from the same
ion. The ratio of transitions such as 
[OIII]~$\lambda 4363$ to [OIII]~$\lambda\lambda 5007,4959$,
indicates the nebular electron temperature ($T_e$), and, along with
the electron density, yields
a direct estimate of the chemical abundance.
Unfortunately, the relatively weak [OIII]~$\lambda 4363$ line
is only observed in the hottest, most metal-poor H~II regions, and 
is not detected in more metal-rich environments. 
To address this issue, H~II region photoionization models
that predict the relative strengths of $\Ha$,  $\Hb$, and 
strong forbidden lines of [OII], [OIII],
[NII], [SII], and [SIII], have been calibrated against 
H~II regions with direct abundance estimates 
\citep{pagel1979,evans1985,kewley2002}. 
With these empirically calibrated models,
it is possible to infer approximate chemical abundances over
a wide range of metallicity,
using only the ratios of
strong, easily detectable H~II region emission lines.

The most commonly-used strong-line abundance indicator is the 
``$R_{23}"$ method, which relates the nebular oxygen
abundance to the ratio of ([OII] + [OIII])/$\Hb$ \citep{pagel1979}. 
The $R_{23}$ method has been employed to determine the metallicities
of individual H~II regions in the local universe, and
the average metallicities of entire star-forming
galaxies in the distant universe \citep{pettini2001,lilly2003}.
The double-valued nature of the $R_{23}$ calibration limits its utility
unless an independent line-ratio can be used to determine whether a galaxy
is on the upper, metal-rich branch or the lower, metal-poor branch. 
Furthermore, in objects for which [OIII]~$\lambda 4363$ or other
temperature-sensitive auroral lines are detected, 
metallicities inferred from $R_{23}$ are
systematically higher by $0.2-0.5$~dex than
those computed directly from the $T_e$ method
\citep{kennicutt2003}.

As presented in section~\ref{sec:results}, we have measured 
[NII]/$\Ha$ from the sample of UV-selected star-forming $z\sim 2$ galaxies. 
This line ratio
can also be used to infer approximate chemical abundances.
Unlike $R_{23}$, the [NII]/$\Ha$ ratio increases monotonically
with increasing metallicity (at least up to solar metallicity).
Another benefit of the [NII]/$\Ha$ ratio is the small difference
in wavelength between the two features, which reduces the 
importance of accurate flux-calibration and dust extinction correction.
The utility of the ``$N2$" index 
$(N2\equiv \log(\mbox{[NII]}\lambda 6584/\Ha) )$ has recently been
revisited empirically by \citet{pp2004}, building on
work by \citet{denicolo2002}. \citet{kewley2002}
demonstrate theoretically how $N2$ depends on metallicity using detailed
photoionization modeling. The $N2$ metallicity dependence
stems from several effects. As the metallicity increases,
the ionizing stellar spectrum softens, 
and the overall ionization
decreases, as demonstrated by the increasing ratio of [OII]/[OIII]. 
As the ratio of [OII]/[OIII] increases, so does the ratio of 
[NII]/[NIII]. Furthermore, at (O/H) metallicities above 
$\sim 0.2\times \mbox{(O/H)}_{\odot}$, the ratio of N/O increases
due to the secondary nature of Nitrogen in this regime \citep{henry2000}.
Limitations of the $N2$ method include the fact that $N2$
is sensitive to the value of the ionization parameter
as well as to metallicity. Also, the N/O ratio can vary as a function
age or star-formation history (not simply abundance). Finally,
\citet{kewley2002} show with detailed photoionization models
that while $N2$ is a monotonically increasing function of metallicity
at less than or equal to solar (O/H), it saturates at solar metallicity
and even turns over towards lower values (at $Z> 2Z_{\odot}$).

Despite these limitations, the $N2$ indicator offers
us rough estimates of the H~II region abundances,
especially using the latest empirical calibration
of \citet{pp2004}. This calibration only includes (O/H)
determinations from the direct $T_e$ method
or detailed photoionization modeling. The linear 
relationship between $\log(\mbox{O/H})$ and
$N2$ thus derived is:

\begin{equation}
12+\log\mbox{(O/H)} = 8.90 + 0.57 \times N2
\label{eq:N2}
\end{equation}
For any individual $N2$ measurement, equation~\ref{eq:N2}
should yield the correct (O/H) abundance to within $\pm 0.18$~dex
(68\% confidence). Using this relationship, and assuming its validity
for interpreting the integrated spectra of high-redshift
star-forming galaxies,
we determine (O/H) abundances for the sample of 
$z\sim 2$ galaxies with NIRSPEC [NII]/$\Ha$ measurements.
The (O/H) abundances and associated uncertainties
are listed in Table~\ref{tab:abun}. For comparison
with other recent work, we also list the (O/H) abundances
implied by the relation presented in \citet{denicolo2002} 
\begin{equation}
12+\log\mbox{(O/H)} = 9.12 + 0.73 \times N2
\label{eq:N2d}
\end{equation}
which are systematically higher than those computed
from the \citet{pp2004} relation. In the remaining
discussion, however, we emphasize the results from the
\citet{pp2004} calibration, since it is
based on a more homogeneous sample of (O/H) abundances.

All of the
galaxies for which we detected [NII] have (O/H)
abundances consistent with the most recent estimates
of the solar abundance, $12+\log(\mbox{O/H})_{\odot} = 8.66 \pm 0.05$,
and that of the Orion nebula, $12+\log(\mbox{O/H})_{Orion} = 8.64 \pm 0.06$
\citep{allende2002,asplund2004,esteban1998}. 
Furthermore, [NII]/$\Ha$ saturates
at $\mbox{(O/H)} \geq \mbox{(O/H)}_{\odot}$
and even turns over towards lower values as
the increased cooling at the highest metallicities
lowers the H~II region
$T_e$ to the point that the collisionally-excited
[NII] emission is reduced \citep{kewley2002}. Since equation~\ref{eq:N2}
is a linear fit, which does not take this turn-over
into account, the highest
[NII]/$\Ha$ measurements in the current sample -- those
of BX599, BX453, BX344, and BX663 -- could very well
correspond to super-solar metallicity values.
To address this uncertainty, \citet{pp2004} propose the use of the ``$O3N2$"
indicator, which consists of the ratio of two line ratios:
([O~III]~$\lambda 5007/\Hb$)/([N~II]~$\lambda 6583/\Ha$). 
While [NII]/$\Ha$
saturates at high metallicity, [OIII]/$\Hb$
continues to decrease such that $O3N2$ also
continues to decrease. The monotonically decreasing behavior
of $O3N2$ with increasing metallicity has been
calibrated from $0.3 \leq Z/Z_{\odot} \leq 2.5 $,
and exhibits significantly less scatter than the corresponding relation
using $N2$ alone.
A clear next step is to obtain $H$-band [OIII]/$\Hb$
spectra for the galaxies in this sample.

In the preceding discussion, we have interpreted the [NII]/$\Ha$ ratios
in terms of the H~II region oxygen abundances that they imply.
This interpretation rests on the assumption that the observed emission
lines originate in H~II regions 
excited by the photoionizing radiation from massive O and B 
stars. The contribution from diffuse ionized gas (DIG)
to the integrated emission-line spectra
of local star-forming galaxies
gas can enhance the [NII]/$\Ha$ ratio by different amounts ranging
from less than 0.1 dex 
to as much as a factor
of two, relative to what is measured in
individual H~II regions \citep{lehnert1994}. The DIG component could potentially
bias our interpretation of [NII]/$\Ha$ ratios towards higher metallicities.
Currently, we have no constraints
on the contribution of DIG to the spectra
presented here, which would require spatially-resolved $\Ha$ narrow-band
imaging of the cosmologically-distant galaxies in our sample. 

Other possible excitation mechanisms include photoionization
by an AGN power-law continuum, or shock-heating, as
seen in low-ionization nuclear emission regions (LINERS) 
and supernova remnants. In the future, combining [NII]/$\Ha$
with measurements of [OIII]/$\Hb$
will clearly discriminate among
the different excitation mechanisms for these galaxies.
For now, there is no evidence pointing to an AGN as the
source of excitation. 
In addition to the NIRSPEC
rest-frame optical spectra, we have also analyzed the
Keck~I/LRIS rest-frame UV spectra of these galaxies.
An AGN manifests itself in the rest-frame UV with strong
Ly$\alpha$ emission, and 
high-ionization emission lines including Si~IV, Ci~IV, and N~V 
\citep{steidel2002}.
Only one of the seven objects in the $K_s$ selected sample, BX663,
displays strong Ly$\alpha$ emission in its rest-frame UV spectrum,
but no other emission lines are detected.
In the remainder of the sample, Ly$\alpha$ is detected in either strong
absorption or with an equivalent width of $\sim 0$~\AA, and 
none of these rest-frame UV spectra show evidence for high-ionization 
emission lines.

A consistency check on the possible AGN contribution is provided 
by the NIRSPEC spectra. Different types
of extragalactic emission-line regions can be
separated according to the relative values of their
emission-line strengths. For example, \citet{bpt1981} and 
\citet{veilleux1987} show that AGN and LINERS 
tend to have higher [NII]/$\Ha$ ratios (i.e. on the
order [NII]/$\Ha \sim 1$) than star-forming galaxies,
and that AGN have higher [OIII]/$\Hb$ ratios 
than star-forming galaxies for a given [NII]/$\Ha$ ratio. 
While we currently do not have $H$-band [OIII]/$\Hb$ spectra for the
$z\sim 2$ sample, another way in which AGN and star-forming galaxies can be
distinguished is according to the [OI]~6300/$\Ha$ ratio. AGN and
shock-heated regions have [OI]~6300/$\Ha \sim 0.1-1$, whereas
star-forming galaxies have [OI]~6300/$\Ha$ ratios of $<0.1$.
Since the detection of [OI] at less than 0.10 times the strength of $\Ha$
pushes the limit of what can be achieved with the individual galaxy
spectra, we construct higher-S/N composite NIRSPEC spectra of the 
entire sample of seven galaxies, and the subset of four of these
with the highest [NII]/$\Ha$ ratios. [OI]~6300 is not
detected in either case, yielding a limit of [OI]~6300/$\Ha < 0.05$.
The lack of significant [OI]~6300 emission in the composite NIRSPEC 
spectra lends further support to the
interpretation of the emission line strengths in the context
of stellar H~II regions.

\section{Discussion}
\label{sec:discussion}

\subsection{The Metallicity-Luminosity Relation}
\label{sec:zmb}

\begin{figure*}
%\plotone{zmb_garnett.eps}
\plotone{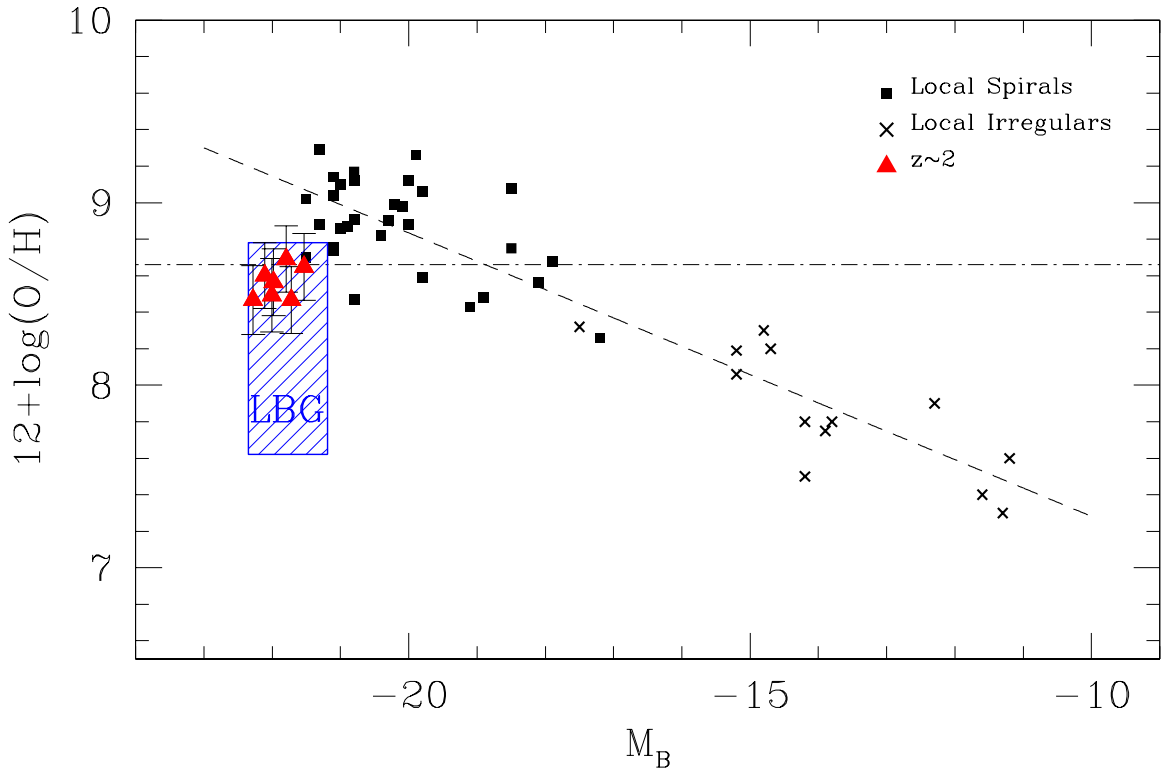}
\caption{The Metallicity-Luminosity relationship.
Data for local spiral (black squares) and irregular (black crosses)
galaxies are taken from \citet{garnett2002}, and display
the well-studied strong correlation between
(O/H) abundance and absolute $B$ luminosity.
The dashed line indicates a least-squares fit to the local
data, while the dot-dashed line indicates solar (O/H) abundance.
$K_s \leq 20.0$ UV-selected $z\sim2$
objects (red triangles) are over-luminous for their (O/H) abundances,
derived using the $N2$ calibration of \citet{pp2004},
but lie closer to the
relationship than $z\sim 3$ LBGs (blue shaded region).
The majority of the $z\sim 2$ data points represent lower limits
in (O/H), since the measured $N2$ values lie in a regime where
this line ratio becomes insensitive to increasing metallicity.
}
\label{fig:zmb}
\end{figure*}

We have presented the rest-frame optical spectra
of a sample of UV-selected star-forming galaxies 
that are also bright in the rest-frame optical. The
[NII]/$\Ha$ line ratios of the majority of these galaxies indicate that they
have already been enriched to at least
solar oxygen abundance at a lookback 
time of $\sim 10$~Gyr. It is worthwhile to consider their metallicities and
luminosities along with those of local star-forming
and elliptical galaxies. Over a factor
of 100 in (O/H) abundance and 11 magnitudes in blue luminosity, nearby
star-forming spiral and irregular galaxies exhibit a strong trend of
higher metallicity with increasing luminosity \citep{garnett2002}.
Elliptical galaxies show a similar correlation between metallicity
and luminosity, although their metallicities are
measured from absorption indices in late-type stars rather than H~II emission
line strengths \citep{brodie1991}. This correlation can be explained
if less luminous (and less massive) galaxies have larger gas fractions,
either because they have lower star-formation rates per unit mass
than more massive galaxies, or because they are younger. Another
way of explaining the correlation is through
galactic-scale outflows, which could remove 
a larger fraction of gas and metals from less massive galaxies,
which have shallower gravitational potentials, thereby reducing 
the effective heavy element yield \citep{garnett2002}.

Both \citet{kobulnicky2000} and \citet{pettini2001}
have shown that $z\sim3$ Lyman Break Galaxies 
(LBGs) are $\sim 2-4$ magnitudes overluminous
for their metallicities, when compared with the local metallicity-luminosity
relation. Despite large uncertainties in LBG metallicities,
the upper bound of the abundances allowed by the observed LBG $R_{23}$
values still falls a factor of three in metallicity below the 
local metallicity--luminosity relationship. In Figure~\ref{fig:zmb}
we plot the local metallicity--$M_B$ relationship (with data
compiled by \citet{garnett2002}) along
with the shaded region allowed for LBGs. In this figure, we 
also include the luminosity and metallicity measurements for the 
rest-frame optically luminous $2.1 < z < 2.5$ star-forming galaxies
presented here.
For both the LBGs and $z\sim 2$ galaxies, 
$M_B$ was calculated using
best-fit model spectral energy distributions (see section~\ref{sec:masses})
to interpolate 
the observed broadband $U_nG{\cal R}K_s$ colors to rest-frame $B$.

While the specific (O/H) values and errors implied by
the linear \citet{pp2004} relationship place the $z\sim 2$
galaxies below the metallicity-luminosity relationship
for local galaxies, it must also be borne in mind that
our estimates of (O/H) represent lower 
limits because of the relative insensitivity of the 
$N2$ index when (O/H) is higher than solar.  
Furthermore, since (O/H) abundances in this plot are derived with
the $R_{23}$ method for the $z\sim 3$ 
and $z=0$ galaxies
and with the $N2$ method for the $K_s\leq 20.0$
$z\sim 2$ galaxies, it is important to acknowledge possible systematic 
differences between the two methods. As mentioned above, these
systematic differences are definitely relevant
in the super-solar metallicity regime
where [NII]/$\Ha$ is no longer sensitive to increasing (O/H) abundance.
The oxygen abundance derived
from [NII]/$\Ha$ may also be $\sim 0.1-0.2$~dex lower than the
abundance derived from $R_{23}$
at lower metallicities, where [NII]/$\Ha$ {\it is} sensitive to 
metallicity \citep{vanzee1998}. 
To estimate (O/H) from the $z\sim 2$ [NII]$\Ha$ ratios
in a manner more consistent with the $R_{23}$ abundance
determinations for local H~II regions,
we examined the emission line data for a set of H~II regions located
in eight galaxies in the \citet{garnett2002} sample.
Uniformly measured $N2$ and $R_{23}$ values for these H~II regions
are contained in \citet{vanzee1998}. 
The $z\sim 2$ galaxies
in our sample have [NII]/$\Ha$ ratios that span from $0.17$ to $0.43$.
H~II regions in the \citet{vanzee1998}
sample with [NII]/$\Ha$ in this range are found to have 
$12+\log\mbox{(O/H)} = 8.50-9.40$, using the upper-branch $R_{23}$
calibration of \citet{zaritsky1994}, which corresponds to
$0.7-5.5 \times \mbox{(O/H)}_{\odot}$. 
While the $R_{23}$ calibration is in itself untested at these very high
values of metallicity and may well suffer from its own systematic
uncertainties, it seems very likely that the $z\sim2$ star-forming galaxies
we have observed in this work are enriched to levels comparable to
those of the most metal-rich and massive spiral and elliptical
galaxies in the local universe.

Whether solar or super-solar, 
the metallicities of these galaxies are likely to grow
further from $z\sim2$ to the present time, since we observe them while they
are still in the process of actively forming stars at rates of 
$24-60 M_{\odot}\mbox{ yr}^{-1}$.
In one possible scenario, proposed
by \citet{maier2004} to explain the range of (O/H) and $M_B$ measurements of
star-forming galaxies at $z\sim 3$, $z\sim 0.5$, and $z\sim 0$,
the rest-frame optically bright $z\sim 2$ galaxies in our
sample will evolve both in metallicity and luminosity by $z\sim 0$; their
oxygen abundances will increase by a factor of two, and their
luminosities will fade by $\sim 2$ magnitudes.
Unfortunately, at this point,
we cannot precisely determine the amount of future star-formation
and enrichment in these galaxies (and therefore tracks in metallicity-luminosity
space as a function of redshift) since we have no information about their 
molecular gas content or future episodes of 
gas accretion and infall. 

In any case, it is important to obtain deeper [NII] and $\Ha$ 
measurements of $K_s\geq 20$ star-forming $z\sim 2$ galaxies
in order to determine the $z\sim 2$ metallicity-luminosity relationship
over a larger range of rest-frame optical luminosities and relate
star-forming $z\sim 2$ galaxies to those at lower redshift.
As a first step, we compare galaxies in the current sample with those
presented in \citet{erb03}, in terms of their
rest-frame optical luminosities (corrected for $\Ha$ emission)
and [NII]/$\Ha$ ratios. 
The current sample of $K_s \leq 20.0$ galaxies
is characterized by $\langle M_{opt}\rangle=-23.29$
and $\langle \mbox{[NII]}/\Ha\rangle=0.27$, corresponding to
$\langle 12+\log\mbox{(O/H)}\rangle=8.58$. 
Of the 17 galaxies with $\Ha$ emission-line flux measurements in
\citet{erb03}, 14 also have measured $K_s$ magnitudes, from which 
rest-frame optical luminosities can be calculated. 
The \citet{erb03} galaxies
with $K_s$ measurements are fainter on average than the $K_s\leq 20.0$
sample, with $\langle M_{opt}\rangle=-22.16$.
It was not possible to estimate 
[NII]/$\Ha$ on an individual basis for most of the \citet{erb03} galaxies,
since [NII] is not significantly detected in the majority of the spectra.
To determine the average [NII]/$\Ha$, we constructed a composite NIRSPEC
spectrum from the 14 galaxies with $K_s$ measurements. In this spectrum,
we found [NII]/$\Ha$=0.10, corresponding
to $12+\log\mbox{(O/H)} =8.33$. Therefore, the galaxies from \citet{erb03} 
are $\sim 1.1$~mag fainter in the rest-frame optical and $0.25$~dex lower in
$\log\mbox{(O/H)}$ than the current sample. This trend marks the preliminary
establishment of a metallicity--luminosity relationship at $z\sim 2$.

\subsection{The Masses of Metal-Rich Star-forming $z\sim 2$ Galaxies}
\label{sec:masses}

\begin{deluxetable*}{lrrcccccc}
\tablewidth{0pt}
\tabletypesize{\footnotesize}
\tablecaption{Star-formation Rates and Stellar Population Parameters\label{tab:sfr}}
%\rotate
\tablehead{
\colhead{} &
\colhead{} &
\colhead{} &
\colhead{Uncorrected} &
\colhead{Uncorrected} &
\colhead{Model} &
\colhead{} &
\colhead{} &
\colhead{} \\
\colhead{} &
\colhead{} &
\colhead{} &
\colhead{SFR$_{\Ha}$\tablenotemark{c}} &
\colhead{SFR$_{UV}$\tablenotemark{d}} &
\colhead{SFR\tablenotemark{e}} &
\colhead{} &
\colhead{Age\tablenotemark{g}} &
\colhead{$M_{star}$\tablenotemark{h}} \\
\colhead{Galaxy} &
\colhead{$z_{\Ha}$\tablenotemark{a}} &
\colhead{$L_{\Ha}$\tablenotemark{b}} &
\colhead{$(M_{\odot} \mbox{yr}^{-1})$} &
\colhead{$(M_{\odot} \mbox{yr}^{-1})$} &
\colhead{$(M_{\odot} \mbox{yr}^{-1})$} &
\colhead{$E(B-V)$\tablenotemark{f}} &
\colhead{(Gyr)} &
\colhead{$(10^{11} M_{\odot})$}
}
\startdata
Q1623-BX274 & 2.4100 & 4.3 & 34 &  28 &  75 & 0.12 & 1.3 & 1.9 \\
Q1623-MD66  & 2.1075 & 6.5 & 51 &  10 &  65 & 0.23 & 0.9 & 0.9 \\
Q1623-BX344 & 2.4224 & 7.9 & 62 &  8  &  49 & 0.20 & 1.6 & 1.9 \\
Q1623-BX453 & 2.1816 & 4.9 & 39 &  16 & 174 & 0.27 & 0.4 & 0.9 \\
Q1623-BX528 & 2.2682 & 3.0 & 24 &  18 &  44 & 0.11 & 1.7 & 1.9 \\
Q1623-BX599 & 2.3304 & 7.6 & 60 &  22 &  49 & 0.10 & 1.3 & 1.3 \\
Q1623-BX663 & 2.4333 & 7.8 & 62 &  12 &  33 & 0.13 & 2.0 & 2.3 \\
\enddata
\tablenotetext{a}{Vacuum heliocentric redshift of H$\alpha$ emission.}
\tablenotetext{b}{$\Ha$ luminosity in units of $10^{42}\mbox{ erg s}^{-1}$.}
\tablenotetext{c}{SFR calculated from $L_{\Ha}$, using the conversion
of \citet{kennicutt1998}.}
\tablenotetext{d}{SFR calculated from $L_{1500}$, probed by the
$G$ apparent magnitude and using the conversion of \citet{kennicutt1998}.}
\tablenotetext{e}{SFR calculated by fitting a \citet{bc2003} $\tau=1$~Gyr
model reddened by dust extinction
to the $U_nG{\cal R}K_s$ colors of the galaxies. The best-fit value
of $E(B-V)$ allows for the calculation of the dust-corrected value
of the star-formation rate, listed here.}
\tablenotetext{f}{The best-fit value of $E(B-V)$,
assuming a \citet{calzetti2000} dust-attenuation
law as a function of wavelength.}
\tablenotetext{g}{The best-fit value of stellar population age associated
with the current episode of star formation,
assuming an exponentially declining star-formation
history with $\tau=1$~Gyr.}
\tablenotetext{h}{Formed stellar mass computed by integrating
the best-fit exponentially-declining SFR between $t=0$ and the
best-fit age.}
\end{deluxetable*}

The high H~II region metallicities of the $K_s$-bright star-forming
$z\sim 2$ galaxies suggest mature systems that have processed
a significant fraction of their baryonic material into
stars. Modeling the broad-band optical and near-infrared $(U_nG{\cal R}K_s)$
spectral energy distributions
of these galaxies with population synthesis codes
provides an independent test of this hypothesis. Following
a similar procedure to that of \citet{shapley2001}, we use \citet{bc2003} models
with solar-metallicity and a Salpeter initial mass function (IMF) 
to deduce stellar masses and ages.
Dust extinction is taken into account with a \citet{calzetti2000}
starburst attenuation law.\footnote{\citet{reddy2004}
have shown that using the Calzetti law to infer unobscured
star-formation rates from UV colors and magnitudes accurately predicts the
average X-ray and radio continuum fluxes of $z\sim 2$ star-forming galaxies.}
In order to determine how well the stellar masses and 
ages can be constrained, we investigated a range 
of star-formation histories of the form $SFR(t) \propto \exp(-t/\tau)$,
with e-folding times of $\tau=0.01, \mbox{ } 0.05, \mbox{ } 0.10,
\mbox{ } 0.2, \mbox{ } 0.5, \mbox{ } 1, \mbox{ } 2, \mbox{ and } 5$~Gyr,
as well as continuous star-formation models. Before modeling the
colors, we corrected the $K_s$ magnitudes for the effects
of $\Ha$ and [NII] nebular emission.\footnote{For a full description
of the modeling procedure, see \citet{shapley2001}.}

Five of the seven galaxies show evidence of having sustained
active star formation over long timescales, are not well described by steeply
declining star-formation histories, and appear
to have formed at least $\sim 10^{11} M_{\odot}$ of stars. 
Only $\tau\geq 200$~Myr models provide acceptable fits to their SEDs, yielding
ages that are significant fractions of the Hubble time at $z\sim 2$.
In fact, while constant-star-formation models provide
statistically excellent fits to the photometry for these five
galaxies, the associated best-fit ages are older than the
ages of the Universe at each galaxy's redshift. To avoid this
contradiction, we only consider
models which yield best-fit ages causally allowed by the finite age
of the universe. The $\tau=200$~Myr models yield stellar masses
of $0.8 - 1.3 \times 10^{11} M_{\odot}$ and ages of $0.6-0.8 $~Gyr,
whereas the maximally-allowed $\tau$ models for each object yield
masses of $1.9-2.8 \times 10^{11} M_{\odot}$ and ages of $2.1-2.5$~Gyr.
Both ground-based $J$ and $H$-band, and {\it Spitzer} 
rest-frame near-infrared photometry will 
help to discriminate between $\tau=200$ and
$\tau \geq 1$~Gyr models.
The colors of the remaining two galaxies, BX453 and MD66, provide very little
restriction on star-formation history. Both of these
galaxies can be fit by steeply-declining or constant star-formation
models. The best-fit masses for BX453 range from $0.4-0.9 \times 10^{11} M_{\odot}$
with associated ages ranging from $0.1 - 0.5$~Gyr, whereas
the masses for MD66 range from $0.5-1.2 \times 10^{11} M_{\odot}$ and
ages from $0.3-1.7$~Gyr. 
While less massive than the majority of
the sample, BX453 and MD66 have already formed
a significant fraction of the stellar mass of an $L^{*}$ galaxy in the
local universe \citep{cole2001}.

\begin{figure*}
%\plotone{fitpanel.eps}
\plotone{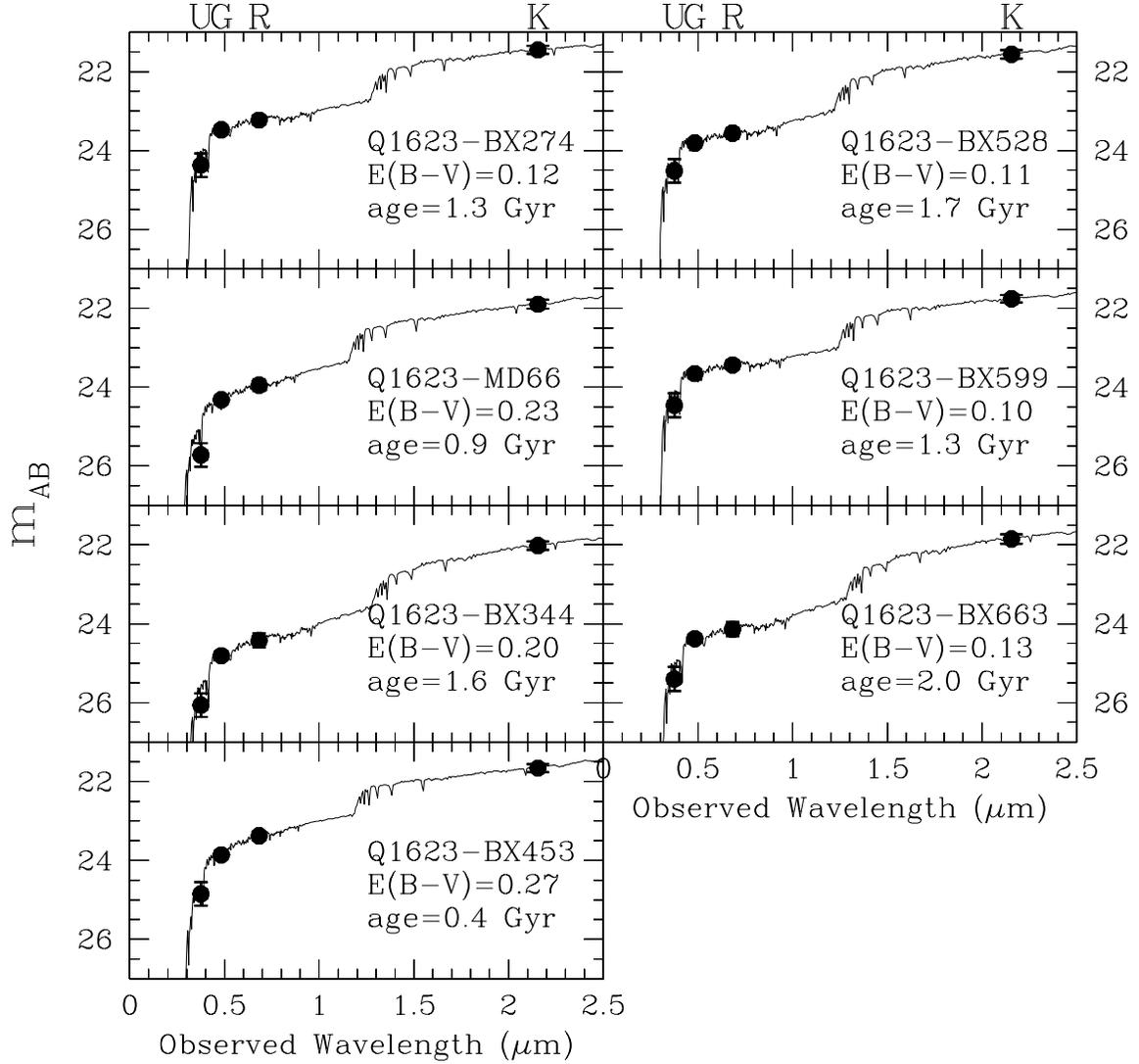}
\caption{
Best-fit \citet{bc2003} models with $\tau=1$~Gyr. These plots correspond
to the stellar population parameters listed in Table~\ref{tab:sfr}.
$U_nGRK_s$ data points are plotted, as are photometric error bars,
which are typically
smaller than the size of the data points. The best-fit models
are all characterized by moderate amounts of dust extinction and
significant Balmer Breaks, indicating
the extended time-scale over which star-formation has proceeded
and $\simgt 10^{11} M_{\odot}$ of stellar mass has built up.
}
\label{fig:models}
\end{figure*}

For all seven galaxies, the
uncertainties in star-formation history lead to systematic
uncertainties in the stellar mass of a factor of $\sim 2$.
Furthermore, if there is an underlying maximally-old population,
hidden at rest-frame UV-to-optical wavelengths
by the current episode of star formation, the stellar masses
derived here could underestimate the total stellar
mass by up to a factor of five \citep{papovich2001}.
Both the inferred dust extinction, $E(B-V)$,
and dust-corrected star-formation rate depend
systematically on the assumed star-formation histories; specifically,
the best-fit models are characterized by more
dust-extinction and higher bolometric star-formation rates when we assume
larger $\tau$ values. Assuming the maximally-allowed
$\tau$ for each galaxy yields $E(B-V)=0.12-0.27$ for the sample,
and $SFR=30-180 M_{\odot}\mbox{ yr}^{-1}$.
In Table~\ref{tab:sfr}, we list the best-fit stellar-population
parameters, assuming $\tau=1$~Gyr, which provides
causally allowed ages for all galaxies in the sample.
Figure~\ref{fig:models} shows plots of the best-fit
$\tau=1$~Gyr models.
While these objects were selected to be at $z\sim 2$
because they display the colors of star-forming
galaxies, it is clear that they are not only metal-rich,
but also have large stellar masses.

The $z\sim 2$ star-forming galaxies in the sample of
\citet{steidel2004} were selected to have 
intrinsic rest-frame UV colors very similar to those of star-forming
LBGs at $z\sim 3$. In relating the two samples of galaxies,
it is important to consider if plausible progenitors
of the massive, metal-rich $K_s \leq 20.0$ ``BX'' galaxies
can be found among the sample of $z\sim 3$ LBGs with near-IR
photometry, whose stellar populations were modeled by \citet{shapley2001}.
To address this question, we re-model the \citet{shapley2001} sample,
using the most current release of the Bruzual \& Charlot
population synthesis codes \citep{bc2003}. 
For a consistent comparison
with the best-fit values presented in Table~\ref{tab:sfr}
we assume exponentially declining star-formation histories, with
$\tau=1$~Gyr. Evolving all the
$z\sim 3$ best-fit models forward in time, 
we compute the star-formation rates and stellar masses formed
by $z=2$. Assuming that $E(B-V)$ does not change between
$z\sim 3$ and $z=2$, we also compute the predicted ${\cal R}$ and $K_s$
magnitudes, and $U_n -G$, $G-{\cal R}$ colors. The predicted UV colors
can be used to determine if a galaxy would be recovered by the $z\sim 2$ UV 
criteria of \citet{adelberger2004}, and the predicted near-IR magnitudes
indicate whether or not a galaxy would be included in the rest-frame
optically bright sample presented here, with $K_s \leq 20.0$.
At least 15\% of the $z\sim 3$ LBGs fit by 
\citet{shapley2001} have stellar population models
time-evolved to $z=2$ with stellar masses, star-formation
rates, $E(B-V)$ and age within the range of properties of the
$K_s \leq 20.0$ BX galaxies presented here.
The time-evolved $U_nG{\cal R}$ colors of these LBGs satisfy the UV criteria
of \citet{adelberger2004}, and they also have $K_s\leq 20.0$.
Therefore, at least 15\% of the
$z \sim 3$ LBGs modeled by \citet{shapley2001} constitute
plausible progenitors for the $K_s$-bright BX objects.
As shown below, the $K_s$-bright BX's
have a comoving number density that's only 14\% as high as
that of LBGs in the \citet{shapley2001} sample, 
so we conclude that most of them could have been LBGs in the past.

It is now possible to estimate the stellar mass density associated
with $K_s \leq 20.0 $ UV-selected $z\sim 2$ objects. First we estimate the 
average mass of the 7 objects with [NII]/$\Ha$ measurements. Taking
into account the systematic uncertainties due to the range of
allowed star-formation histories, we find 
$\langle M_{star} \rangle = 1.4 \pm 0.5 \times 10 ^{11} M_{\odot}$. This
average mass should be representative of the entire sample of $K_s \leq 20.0$
$z\sim 2$ galaxies since the distributions of 
$G-{\cal R}$ and ${\cal R}-K_s$ colors 
of the seven galaxies with mass estimates are similar to 
those of the entire sample of rest-frame optically luminous objects.

\begin{deluxetable*}{rcc}
\tablewidth{0pt}
\tabletypesize{\footnotesize}
\tablecaption{Fraction of $2.0 \leq z \leq 2.7$ BX objects vs. ${\cal R}$ magnitude\label{tab:rzbx}}
%\rotate
\tablehead{
\colhead{${\cal R}$ Magnitude Range} &
\colhead{Fraction of BX objects with $2.0 \leq z \leq 2.7$\tablenotemark{a}} &
\colhead{Q1623 BX objects with $K_s \leq 20.0$ and no z}
}
\startdata
${\cal R} \leq 21.7$\tablenotemark{b} & 0/48 & 39 \\
$21.7 < {\cal R} \leq 23.0$   & 6/51 & 13 \\
$23.0 < {\cal R} \leq 24.0$ & 145/301  & 7 \\
$24.0 < {\cal R} \leq 25.5$  & 386/556  & 8 \\
\enddata
\tablenotetext{a}{These statistics are based on the entire sample
of 956 spectroscopically identified BX objects. }
\tablenotetext{b}{We use ${\cal R}=21.7$ as a cut-off since
all spectroscopically-confirmed BX objects at
brighter magnitudes are stars and QSOs.}
\end{deluxetable*}

Next we estimate the comoving space density of $K_s \leq 20.0$
$z\sim 2$ UV-selected galaxies. In the Q1623 WIRC pointing, which
covers $92\mbox{ arcmin}^{2}$, there
are 87 ``BX'' objects with $K_s\leq 20.0$, including 13 galaxies with
confirmed redshifts
between $2.0 \leq z \leq 2.7$, one broad-lined QSO
at $z=2.53$, four stars, and two
low-redshift galaxies at $z\sim 0.2$.
The remaining 67 photometric $z\sim 2$ candidates
have either not been attempted
spectroscopically or have been attempted unsuccessfully,
with undetermined redshifts. 
To estimate the number of $K_s\leq 20.0$ galaxies without
spectroscopic redshifts
that are likely to fall in the redshift range of interest, 
we use the redshift statistics as a function of ${\cal R}$
magnitude from our total sample of 956 BX objects
with spectroscopic information \citep{steidel2004}. 
Based on the BX high-redshift fractions and the
distribution of ${\cal R}$ magnitudes for the 
67 $K_s \leq 20.0$ sources without redshifts
presented in Table~\ref{tab:rzbx} (i.e. multiplying 
the second column of Table~\ref{tab:rzbx} by the third column), 
we estimate that an additional 11 $K_s \leq 20.0$ galaxies without
measured redshifts are likely to be at $2.0 \leq z \leq 2.7$.
The total number
of $K_s \leq 20.0$ BXs at $2.0 \leq z \leq 2.7$ is therefore $N=13+11=24$.
Weighted by the BX redshift selection function 
\citep{steidel2004,adelberger2004}, the comoving volume 
between $2.0 \leq z \leq 2.7$ is $1496 \mbox{ Mpc}^3\mbox{arcmin}^{-2}$. 
Over the entire Q1623 WIRC pointing,
the space-density of $2.0 \leq z \leq 2.7$ UV-selected objects with
$K_s \leq 20.0$ is therefore 
$n=1.7 \times 10^{-4} \mbox{ Mpc}^{-3}$. 
The space density of massive galaxies is of course
prone to large sample variance because such objects are expected 
to be strongly clustered. While not yet properly quantified, 
initial statistics from our two other fields with $K_s$-band
data suggest variations of a factor of $\sim 2$ from field to field.

Multiplying this space
density by the average mass of the rest-frame optically luminous
objects we obtain 
$\rho = n \times \langle M_{star} \rangle = 2.5 \pm 0.9 \times 10^{7} M_{\odot}\mbox{ Mpc}^{-3}$, which is 4\% of the $z=0$ stellar mass density
determined by \citet{cole2001}. The $K_s\leq 20.0$ objects represent only
$\sim 10$\% of the sample of BX objects at $z\sim 2$, and since we expect to find
massive systems down to significantly fainter $K_s$ magnitudes, this figure
represents an extreme lower limit on the stellar mass density associated
with UV-selected star-forming galaxies at $z\sim 2$. In the future,
when we have the distribution
of stellar masses over a much wider range in luminosity in hand, we will
determine a more robust constraint on the stellar mass density at $z\sim2$,
extending the work of \citet{dickinson2003} to a larger cosmic volume.

\subsection{Comparison with Other Surveys}
\label{sec:surveys}
There are other current surveys targeting massive galaxies in the same redshift range.
It is instructive to compare the properties of the $K_s \leq 20.0$
UV-selected objects presented here with those of the galaxies in the K20 survey
\citep{cimatti2002} and the Gemini Deep Deep Survey \citep[GDDS;][]{abraham2004}.
The K20 project is a spectroscopic
survey of $\sim 500$ $K_s \leq 20.0$ objects, 9 of which have been spectroscopically
confirmed in the redshift range $1.7 < z \leq  2.25$. \citet{daddi2004} demonstrate
that this subsample of galaxies is characterized by dust-corrected star-formation rates of 
$\sim 100-500 M_{\odot}\mbox{ yr}^{-1}$, constant star-formation ages of $0.25-1.7$~Gyr,
and stellar masses of $0.3-5.5 \times 10^{11} M_{\odot}$. Six of the galaxies in
\citet{daddi2004} are at $z>2$, and four of them have optical colors that
satisfy the ``BX'' criteria presented in \citet{adelberger2004}. These four galaxies
have $E(B-V)=0.3$ and constant star-formation ages of $0.7-1.7$~Gyr, very similar
to the properties of BX453 and MD66 when constant star-formation histories
are assumed. Of the three galaxies in \citet{daddi2004} that
are at $z < 2$, two would be recovered by the ``BM'' criteria of \citet{adelberger2004},
which target star-forming $1.5 \leq z \leq 2.0$ galaxies. Since the K20 survey
uses a different optical filter set from that of the \citet{steidel2004}
survey, we calculated the $U_nG{\cal R}$ colors of the
\citet{daddi2004} galaxies by passing the best-fit
redshifted, dust-reddened, constant star-formation models (from their Table 1) through the
$U_nG{\cal R}$ filter set. There appears to be significant overlap between the
$K_s \leq 20.0$ UV-selected objects and the high-redshift tail of the K20 survey.
Furthermore, the space density of $1.7 < z \leq 2.25$ K20 objects is 
$1.6 ^{+1.5}_{-0.8}\times 10^{-4} \mbox{ Mpc}^{-3}$, which is very comparable to
the space density of $K_s \leq 20.0$ UV-selected objects at $2.0 \leq z \leq 2.7$ 
in the Q1623 field,
$n= 1.7 \times 10^{-4} \mbox{ Mpc}^{-3}$. Finally, the distribution of stellar masses
of the K20 and rest-frame optically luminous UV-selected
objects are very similar (with the
exception of one K20 object that has $5.5 \times 10^{11} M_{\odot}$ and would not be
selected as a BX). The additional UV-selection criteria appear
to recover the majority of K20 objects, because most $\geq 10^{11} M_{\odot}$ 
galaxies harbor active star formation at $z\sim 2$. 

While the majority of the 300 galaxies in the GDDS survey are at $z <1.5$,
this infrared-selected $K < 20.6$ survey contains a small high-redshift
tail of $1.6 \leq  z \leq 2.0$ galaxies \citep{glazebrook2004}. Using GDDS galaxies
to trace the evolution of stellar mass-density in the universe,
\citet{glazebrook2004} find a $1.6 \leq z \leq 2.0$ mass density
for $M> M_{lim}=6.3 \times 10^{10} M_{\odot}$ objects of 
$3.3\pm 1.2\times 10^{7} M_{\odot} \mbox{ Mpc}^{-3}$. Below $M_{lim}$,
the $K<20.6$ GDDS becomes incomplete for objects with the highest 
mass-to-light ratios.
The GDDS masses and mass densities were calculated assuming an IMF that
turns over below $0.5 M_{\odot}$ \citep{bg2003,kroupa2001}; it is therefore 
necessary to multiply the GDDS numbers by a factor of 1.82
for a meaningful comparison with the results presented here, 
derived using a Salpeter IMF.
The range of masses of the spectroscopically confirmed GDDS $1.6 \leq z \leq 2.0$ galaxies
is very similar to that of the $K_s \leq 20.0$ UV-selected objects, while the
mass density, computed down to a limiting magnitude of $K < 20.6$
is $\sim 3$ times higher. A more fair comparison in mass
densities will result
when we have determined the stellar masses of fainter UV-selected objects.
Nonetheless, it is still currently possible to 
determine how complementary the UV-selected and
GDDS surveys are in the $z>1.6$ redshift range,
since one of the GDDS survey areas, the SSA22 field, has also been
imaged in the $U_nG{\cal R}$ filter set as part of the $z\sim 3$ LBG
survey \citep{steidel2003}. In the SSA22a field, there are three 
spectroscopically confirmed GDDS galaxies with $1.6 \leq z \leq 2.0$, and  
four with uncertain redshifts of $2.0 < z < 2.2$ \citep{abraham2004}. All three of the $1.6 \leq z \leq 2.0$
galaxies are recovered with the BX/BM $U_nG{\cal R}$-selection criteria, and
three of the four putative $z>2$ objects would be selected as BXs. At $1.4 \leq z \leq 1.6$,
there are 12 GDDS objects in the SSA22a field, half of which would be recovered
by the BX/BM UV-selection criteria. The 50\% overlap between
GDDS galaxies at $1.4 \leq z \leq 1.6$ and BX/BM objects is
not especially meaningful since the BM color criteria are
intended to select galaxies at $1.5 \leq z \leq 2.0$, 
and the BX color criteria
target galaxies at $2.0 \leq z \leq 2.5$--i.e.
$1.4 \leq z \leq 1.6$ lies at the edge of the BX/BM selection
function, where the selection efficiency drops off.
However, the large overlap of the GDDS and BX/BM surveys
at $z>1.6$ is again consistent with the idea that even
$K_s$-band selected surveys that are supposed to be ``mass-selected'' 
include a significant fraction
of star-forming galaxies at $z\sim 2$.

\begin{deluxetable*}{lcccccc}
\tablewidth{0pt}
\tabletypesize{\footnotesize}
\tablecaption{Comparison of Dynamical and Stellar Masses\label{tab:mcomp}}
%\rotate
\tablehead{
\colhead{} &
\colhead{$\sigma$\tablenotemark{a}} &
\colhead{$M_{dyn}$\tablenotemark{b}} &
\colhead{$\tau_{min}$\tablenotemark{c}} &
\colhead{$M_{star}(\tau_{min})$\tablenotemark{d}} &
\colhead{$\tau_{max}$\tablenotemark{e}} &
\colhead{$M_{star}(\tau_{max})$\tablenotemark{f}} \\
\colhead{Galaxy} &
\colhead{(\kms)} &
\colhead{$(10^{11} M_{\odot})$} &
\colhead{(Gyr)} &
\colhead{$(10^{11} M_{\odot})$} &
\colhead{(Gyr)} &
\colhead{$(10^{11} M_{\odot})$}
}
\startdata
Q1623-BX274 & 121 & 0.28 & 0.20 & 1.27 (0.70) & 5.00 & 2.76 (1.52) \\
Q1623-MD66  & 120 & 0.28 & 0.05 & 0.49 (0.27) & $\infty$ & 1.19 (0.65) \\
Q1623-BX344 & 92  & 0.16 & 0.20 & 1.06 (0.58) & 2.00 & 2.18 (1.20) \\
Q1623-BX453 & 61  & 0.07 & 0.01 & 0.49 (0.27) & $\infty$ & 0.92 (0.51) \\
Q1623-BX528 & 142 & 0.58 & 0.20 & 1.09 (0.60) & 2.00 & 2.47 (1.36) \\
Q1623-BX599 & 162 & 0.50 & 0.20 & 0.83 (0.46) & 5.00 & 1.69 (0.93) \\
Q1623-BX663 & 132 & 0.45 & 0.20 & 1.06 (0.58) & 1.00 & 2.35 (1.29) \\
\enddata
\tablenotetext{a}{$\Ha$ velocity dispersion.}
\tablenotetext{b}{Mass calculated from the H$\alpha$ velocity dispersion.}
\tablenotetext{c}{Minimum time-constant of models of the form
$SFR(t) \propto \exp(-t/\tau)$ that provide statistically acceptable
fits to the galaxy colors.}
\tablenotetext{d}{Mass of best-fit model, assuming $\tau=\tau_{min}$. Values
not in parentheses assume a Salpeter IMF extending down to $0.1 M_{\odot}$.
Values in parentheses assume the more realistic \citet{bg2003} IMF,
which has a break at $1M_{\odot}$.}
\tablenotetext{e}{Maximum time-constant of models for which
the best-fit age is younger than the age of the universe. BX453
and MD66 can be fit by $\tau=\infty$, i.e. constant star-formation models, while the remaining
galaxies' best-fit ages exceed that of the universe when $\tau=\infty$.}
\tablenotetext{f}{Mass of best-fit model, assuming $\tau=\tau_{max}$.
Values with and without parentheses have the same meaning as defined in
note (d).}
\end{deluxetable*}

\subsection{Stellar vs. Dynamical Masses}
\label{sec:mstarmvir}
The preceding discussion relies on population synthesis models
of the galaxy colors to infer stellar masses and therefore
the comoving stellar mass density in large galaxies at $z\sim 2$.
Rest-frame optical spectroscopy allows for a completely independent
probe of galaxy mass: the $\Ha$ velocity dispersion. A physical scale-length
and assumption of the geometric configuration are necessary to 
convert the velocity
dispersion into a dynamical mass. In previous determinations of high-redshift
star-forming galaxy dynamical masses \citep{pettini2001,erb03,erb04},
a spherically-symmetric distribution of matter was assumed, and  
the physical scale-length adopted was the typical $z\sim 3$ or $z\sim 2$
galaxy continuum half-light radius, as determined from {\it Hubble Space Telescope}
images, $r_{1/2}=$0\secpoint2--0\secpoint3. From the excellent seeing
conditions during our NIRSPEC exposures, we also determine $\Ha$ half-light
radii of 0\secpoint2-0\secpoint3. Using these radii, and assuming
a spherical geometry, we determine dynamical masses 
with the formula: $M_{dyn}=5\sigma^2 r_{1/2}/G$. 
Dynamical mass estimates are listed
in Table~\ref{tab:mcomp}, as well as stellar masses
inferred from the smallest $\tau$ allowed by the galaxy
photometry, and the largest $\tau$ allowed by the age
of the universe constraint. As discussed in section~\ref{sec:masses},
the stellar mass of the best-fit model is an increasing function of the assumed 
time-constant, $\tau$. The actual stellar mass for each object
lies somewhere between the limits listed in Table~\ref{tab:mcomp}
(neglecting the mass of an underlying maximally-old burst).

For every galaxy, if a Salpeter IMF extending down to 
$0.1 M_{\odot}$ is assumed, the allowed range of stellar mass exceeds the
calculated dynamical mass within $r_{1/2}$.
For BX663, MD66, BX528, and BX599,
this discrepancy is roughly a factor of $\sim 2-4$, whereas for BX344, 
BX274, and BX453, the stellar mass estimate exceeds the dynamical mass
by an order of magnitude. Perhaps this discrepancy is not surprising,
given that the galaxy stellar mass was determined from luminosities
and colors extending over a physical region in some cases
several times as large in radius as 
the region enclosed by the half-light radius. A larger radius
may be more suitable for inferring the dynamical mass.
Indeed, with the assumption of a constant $M/L$ ratio
as a function of radius, the mass enclosed within $r_{1/2}$
should be multiplied by 2 in order to compare with the total stellar mass.

Furthermore, there are observational effects 
that may apply if the geometry is not the 
simple spherically-symmetric one assumed for the calculation.
If the emission is extended along a specific axis,
the possible misalignment of the slit with this axis must be taken
into account. If the configuration of H~II regions is more disk-like,
then inclination effects may also cause us to underestimate the
the circular velocity, and therefore, velocity dispersion. 
\citet{erb03} estimate that the combination
of these two effects may result in an underestimate of a factor of
$\sim 2.5$ in circular velocity. Of course, the numerical coefficient
multiplying $\sigma^2 r$ depends on the assumed geometry as well,
and is unity for a disk geometry (as opposed to 5, assumed
for the spherical configuration). Independent of slit misalignment
and disk inclination effects, the emission within the
$\Ha$ half-light radius may not sample the full range of rotational
velocities of the underlying disk. 
Using local starburst galaxies with well-determined
rotation curves, \citet{lehnert1996}
have shown that the widths of optical emission-lines originating in 
the bright central regions of local
starburst galaxies only sample the solid-body part of the
rotation curve, and therefore do not reflect the
full circular velocity attained on the flat
part of the rotation curve at larger radii. Both
\citet{pettini2001} and \cite{erb03} have pointed
out this limitation in interpreting the rest-frame optical linewidths of
$z\sim 3$ and $z \sim 2$ galaxies in terms of masses.
Finally, stellar masses were determined assuming a Salpeter
IMF extending down to $0.1 M_{\odot}$, though we have
no observational constraints on the low-mass regime of the 
IMF at high redshift. 
An IMF which turns over below $0.5 M_{\odot}$, as assumed
by \citet{glazebrook2004}, would lower the stellar mass estimates
by a factor of $\sim 2$. We assumed the standard Salpeter form
in section~\ref{sec:masses} for the purposes of 
comparing with earlier studies of the stellar
mass content of galaxies at low and high redshift
\citep{cole2001,shapley2001,papovich2001},
though a more realistic form for the low-mass end of the
IMF would significantly reduce the systematic 
discrepancy between stellar and dynamical masses (as shown 
by the values in parentheses in Table~\ref{tab:mcomp}).

Most of the above systematic effects would have 
caused us either to underestimate the dynamical mass, or overestimate
the stellar mass. There is enough uncertainty in our estimates 
of the low-mass end of the IMF, the appropriate radius at which 
to evaluate $M_{dyn}$, and the geometrical configuration of H~II
regions, that the current discrepancy between stellar and dynamical
mass estimates does not represent a crisis. However,
the method of \citet{pettini2001} and \citet{erb03}, using
simple assumptions to compute dynamical
masses, clearly does not account for the amount of stellar mass derived from
modeling the colors of these galaxies.
We must understand the nature of these discrepancies
before we can trace the evolution of mass in
galaxies as a function of redshift.

\section{Summary}
\label{sec:summary}

We have presented the rest-frame optical spectra of a sample of
seven UV-selected
$z\sim 2$ galaxies that also have $K_s\leq 20.0$. We reach the following 
principal conclusions:

1. The majority of these galaxies exhibit high [NII]/$\Ha$ ratios
indicative of solar, and possibly super-solar, metallicities,
at a lookback time of $\sim 10.5$~Gyr.

2. The calibration of \citet{pp2004}, which converts [NII]/$\Ha$
ratios into (O/H) abundances, indicates that
the galaxies in the sample are over-luminous for their metallicities
relative to the local metallicity-luminosity relationship.
However, we may have underestimated the (O/H)
abundances due to the insensitivity of the $N2$ indicator
at super-solar metallicity. In any case,
given the active star-formation rates inferred from
$\Ha$ luminosities and population-synthesis modeling,
these objects will likely become more metal-rich at later times, and are
probably the progenitors of massive spiral and elliptical
galaxies in the local universe. The sample
of $z\sim 2$ galaxies presented in \citet{erb03}
has an average rest-frame optical luminosity $\sim 1.1$~mag
fainter and average [NII]/$\Ha$ ratio three times smaller, relative to
the corresponding values for the current sample,
consistent with the existence of a metallicity--luminosity relationship
at $z\sim 2$.

3. Population-synthesis models of their broad-band 
rest-UV/optical spectral energy distributions suggest
that these galaxies have been forming stars over timescales
comparable to a Hubble time, contain $\geq 10^{11} M_{\odot}$
of stars, and are still actively forming stars.

4. Comparison with surveys such as the K20 and GDDS projects
indicates a significant overlap between near-infrared-
and UV-selected objects. A large fraction of the most massive
galaxies at $z\sim 2$ are actively forming stars.

5. Dynamical mass estimates derived from $\Ha$ velocity dispersions
and galaxy half-light radii are systematically lower than
stellar mass estimates deduced from the broad-band SEDs of
the galaxies presented here. While
several plausible effects can be invoked to explain the difference
in mass estimates, this discrepancy must be understood
if we are to trace the evolution of galaxy masses in the
universe.

\bigskip
We would like to thank Naveen Reddy and Matthew Hunt for their
contributions to the optical imaging and spectroscopic
observations that form the backbone
of our survey of $z\sim 2$ star-forming galaxies. We appreciate
the assistance of an anonymous referee, whose comments improved the manuscript.
We also acknowledge extremely helpful conversations with Don Garnett
and Lisa Kewley.
We wish to extend special thanks to those of Hawaiian ancestry on
whose sacred mountain we are privileged to be guests. Without their generous
hospitality, most of the observations presented herein would not
have been possible. Finally, we thank the staff
at the W.~M. Keck Observatory for their 
assistance with the NIRSPEC observations,
especially Joel Aycock, Gary Puniwai, and Bob Goodrich. 
CCS and DKE have been supported by grants
AST00-70773 and AST03-07263 from the U.S. National Science 
Foundation and by the David and Lucile
Packard Foundation. AES acknowledges support from
the Miller Institute for Basic Research in Science.

%\bibliographystyle{apj}
%\bibliography{apj-jour,lbgrefs}

%\include{table1}
%\include{table2}
%\include{table3}
%\include{table4}
%\include{table5}

\clearpage

\end{document}